# Spectral selectivity from resonant-coupling in microgap-TPV

A. Meulenberg[a] and K. P. Sinha[b]

[a] HiPi Consulting, New Market, MD, 31774, USA (email: mules333@gmail.com)
[b] Department of Physics, IISc, Bangalore 560012, India (email: kpsinha@gmail.com

Near-field energy coupling between two surfaces may arise from frustrated total-internal-reflectance and from atomic dipole-dipole interaction. Such an exchange of energy, if at resonance, greatly enhances the radiation transfer between an emitter and a photovoltaic converter. Computational modeling of selected, but realizable, emitter and detector structures and materials shows the benefits of both near-field and resonance coupling (e.g., with ~ 0.1 µm gaps). In one sense, this is almost an engineering paper. A strong computational model (based on physically-proven concepts and incorporating known and predicted high-temperature properties of acceptable emitter materials) is used to demonstrate the potential of materials (properly-selected to overcome natural limitations) and of structures (carefully crafted to push the limits of present technology) for breaking barriers of thermal conversion at lower-emitter temperatures (< 1000$^{o}$C).

## INTRODUCTION

Interest in ThermoPhotoVoltaics (TPV), which converts light (more accurately thermally-generated, long-wavelength-infrared radiation) from a heated surface into electricity, is beginning to grow as a result of new materials capabilities. Specifically, new semiconductor materials and structures allow photoconverters to convert long-wavelength light into electrical power more efficiently by providing a narrower electrical band gap, better generation of photo-excited minority carriers and their collection at the p-n junction, and reduced recombination dark-current, which controls the open-circuit voltage of the devices. However, there are some approaches, based on improved technical capabilities, which can compound all of these improvements and, potentially, allow TPV to become even more useful.

A requirement of a TPV system is a thermal difference between the emitter and photovoltaic (PV) device. This means that thermal isolation, in the form of distance (or of a vacuum, if the distance is small), must separate the devices. However, it is known that elimination of the gap can greatly enhance the transfer of radiation between them. The ability to increase the transfer of optical photons across a very small gap (a fraction of a wavelength) without allowing heat flow, via phonons, has been demonstrated.[1,2] A theoretical basis for the enhanced optical transfer has even been laid in a quantum mechanical framework.[3]

A number of years ago, following high-temperature (>900C) IV measurements on small (2x2mm) individual InAs photocells and 4-element arrays,[2] it was demonstrated at Draper Labs (but not published) that, even for practical-size devices (2x2 cm) at high temperatures (>700C), the optical throughput can be much-enhanced with the use of a very-narrow vacuum gaps (~ 0.2 μm). This enhancement, through the use of a "submicron" gap, is the basis for Microgap ThermoPhotoVoltaics (MTPV).[4] Despite published confirmation of the theoretical concept, and proof of its technical feasibility, little work has progressed in the implementation of this technology other than that conducted by a startup company (MTPV Corp.),[5] which grew out of this foundational work. We include, herein, calculated results of the remarkable predictions for near-term realizable products.

One enhancement mechanism, available to MTPV and represented in Ref. 1, is purely a physical-optics effect. It involves the development of an "effective" refractive index ($n_g$) within the microgap.[6] As the gap approaches zero width, $n_g$ approaches that of the emitter $n_e$ and PV device $n_{pv}$ (assume $n_e = n_{pv}$ as a form of resonance for optimum results in MTPV). As the difference between refractive indices of the devices and that of the gap diminishes, the critical angle of total internal reflection (TIR) increases and the percentage of thermally-generated light that can escape from the emitter increases. As the gap width, d, approaches zero, the enhancement from this mechanism approaches a maximum of $n_e^2$. This effect (called the $n^2$ effect) is a macroscopic phenomenon and is independent of the distance between <u>individual</u> atomic radiators (excited atoms or dipoles) in the emitter and absorbers (ground-state atoms) in the PV device. The maximum possible improvement from the $n^2$ effect is relative to what is possible from a true black body emitter and a detector with a tuned anti-reflecting (AR) coating). As a consequence, the normal presentation of enhancement with reduced gap, the power transfer of the same materials at sub-micron gaps to that at a large distance, could actually exceed $n^2$. As a practical matter, this could be an advantage when considering methods of rapidly "throttling" electrical power from a thermal source.

The $n^2$ effect depends only on the materials of, and the gap between, the emitter and detector. However, included in the calculations of this paper is enhancement of the optical coupling between atoms that <u>is</u> dependent on the distance between <u>individual</u> radiators and absorbers. This process will be called "resonance-enhancement." ("Proximity-enhancement," shown in Fig. 1, includes both the macroscopic $n_e^2$ effect and the atomic-level resonance-enhancement effect.)

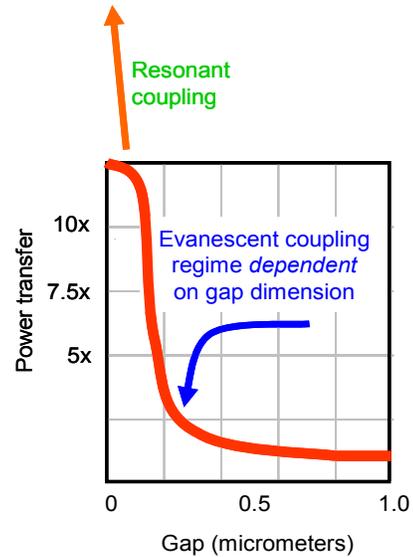

*Fig. 1. Microgap proximity-enhancement modes. The $n^2$ effect (the evanescent coupling regime) is shown for n = 3.55. The pictured resonance-coupling effect (also shown) is specifically vague since it depends on too many variables to be more than representative.*

While somewhat frequency dependent, most of the $n^2$ effect (for TPV applications) will have been achieved by reducing the gapwidth to < 0.1 micrometers (100 nm). In this region, the resonance effects can become identifiable. These gap dimensions, while not producing the spectacular results proposed for 10 nm gap devices, appeared practicable in useful device sizes under development at the time (2003) and were therefore selected for optimization and demonstration. Since surface proximity of the source and detector is the strongest effect with presently-attainable gap dimensions and structures, we feel that planar-geometry techniques (e.g., selective-emission and surface-coupling films) are still likely to be the most productive in the foreseeable future. They are also the easiest to model. The ability to combine (and model) multiple effects and materials in the surface layers is critical to the success of a structure. On the other hand, some of these same techniques can be used in improving thermal emitters to be used in far-field applications. In this latter application, surface coupling is not important and some of the material and structural requirements can be relaxed. Different techniques might then be optimal.

Resonance enhancement, a "beyond $n_e^2$ effect," involves the interaction of dipole (or other) oscillators that are close enough together to be directly "coupled" by another oscillator, the photon[7] or evanescent waves. The basic oscillators may be structures/materials involving electrons, photons, plasmas, or other mechanisms. The resonances may be internal (such as in quantum wells, plasmas, etc.), with adjacent material or structures, with the gap, and/or with the material or structures across the gap. This "correlated" interaction provides enhanced optical coupling and may be compared with the non-correlated interaction of independent creation and absorption of photons. Closer dipoles (in space and in transition energy or frequency) give stronger coupling. Thus, the microgap-enhancement phenomenon is both a proximity effect and a resonance effect.

The computational results below demonstrate the three effects, namely, the gapwidth (source/absorber-separation) dependence, the $n^2$ enhancement, and the resonance enhancement. They will also indicate the effect of PV detector used for the MTPV system. Practical considerations, in terms of large-area microgap implementation, presently limit operation to the $n^2$-effect region. In the future, this limitation may be overcome and the beyond-$n^2$ effects may become dominant. However, it may be possible to "tune" the materials (emitter <u>and</u> detector) so that compound-resonance enhancement may become significant even in this presently-practical gap range.

## CONCEPTS AND PRACTICAL LIMITATIONS

The concepts of near-field enhancement of the useable radiant energy are relatively straight forward.[8] The actual implementation is not so. Typical photovoltaic (PV) devices are designed to convert a solar spectrum (black-body spectrum of >5500K) into electrical power. The non-usable portion of this spectrum (mostly below the semiconductor bandgap energy) is generally less than 10% of the total incident energy. Even that non-useful light is rejected, where economically feasible, because any that is absorbed becomes additional heat, which reduces PV-cell efficiency, and must be removed. For photovoltaic conversion of radiant energy from a thermal emitter (e.g., black-body spectrum of <1300K), the useable energy might be less than 25% of the total radiated. Furthermore, the narrow-bandgap PV devices, required to convert this much of the available radiant energy into electrical energy, are more sensitive to heat than are solar PV cells. Therefore, concentration of the light, which would normally raise solar-cell efficiency, may not be useful for the thermal-photovoltaic (TPV) case with its lower emitter-temperature spectrum. (With cell cooling, the cell efficiency can increase; but, the overall thermal-to-electrical conversion efficiency still goes down.) For these reasons, until recent years, the TPV cell conversion efficiency had been on the order of 2-4%, rather than the 15-23% of commercial solar-PV cells.

### ThermoPhotoVoltaic Cells

The recent demonstration of high-efficiency TPV cells (>20%),[9] along with results on selective filters[10] for them and the work here on selective emission,[11] have been critical parallel developments for TPV work. However, the requirements for MTPV cells and emitters are much more stringent. Heavily-doped semiconductor material (required for ohmic contacts, lateral conductivity between collection grids, and tunnel junctions in high-efficiency Tandem cells[12]) become regions of high free-carrier absorption. For the emitter, this may not be a problem. For the cell, it is a disaster. Thus, while even very thin layers of such material may result in only a few percent efficiency loss in TPV cells, they can create many times that loss in MTPV cells. As a result, special cell structures must be designed for MTPV operation.

Figure 2 pictures a cell designed to cope with these requirements and the important points addressed in its design.

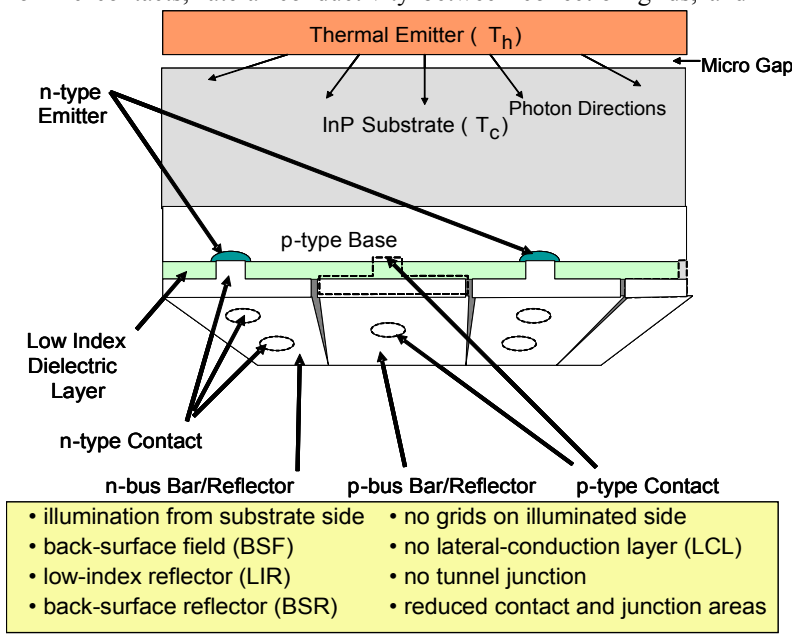

*Fig. 2 An all-back, dot-contact, PV-cell structure shown with design requirements for efficient MTPV operation.*

- illumination from substrate side
- back-surface field (BSF)
- low-index reflector (LIR)
- back-surface reflector (BSR)
- no grids on illuminated side
- no lateral-conduction layer (LCL)
- no tunnel junction
- reduced contact and junction areas

This dot-contact cell is realizable. It is based on long-standard solar-cell technologies[13] and on the design and performance of recent high-efficiency TPV cells that had been demonstrated at the time (2004). It must forgo the tandem-cell configuration[12] (designed to improve TPV-cell efficiency) because of the severe penalty for MTPV operation imposed by the heavily-doped, ohmic-contact layers involved.

Figure 3 is a comparison of optical-absorption characteristics representative of a known >20% efficient TPV cell (developed at Bettis Labs)[9] and the proposed dot-contact cell (Fig. 2), both in an MTPV operating mode (with the effects as described below). A black-body spectrum for the same temperature is displayed along with the total absorption and the absorption in various layers of the cell (with representative values of thickness and doping). The actual emitter spectrum used in the calculation is for a grey body (black-body spectrum, but with emissivity less than 1 and no spectral selectivity) with refractive index of n=3.4. Even with the exceptional care in design and construction of this Bettis-type cell, the area under the curve for total-power-absorbed in the cell beyond the bandedge (>2.2 μm) is comparable to that same area below the bandedge. Thus, though the power output of the cell at a 0.1 μm gap may be >5x that for the same cell and emitter with a large gap, the absorbed long-wavelength light cuts the cell efficiency in half. (Under TPV conditions, this useless absorbed energy is much less than that pictured, e.g., only about 1/10 that of the useable energy absorbed.) It is seen that the most absorbing regions in the TPV cell are the heavily-doped InGaAs layers (see details in Appendix A). By elimination of these layers (except at the dot contacts, ~3% of cell area), the waste-energy absorption of the MTPV cell is reduced by a factor of 6–10x.

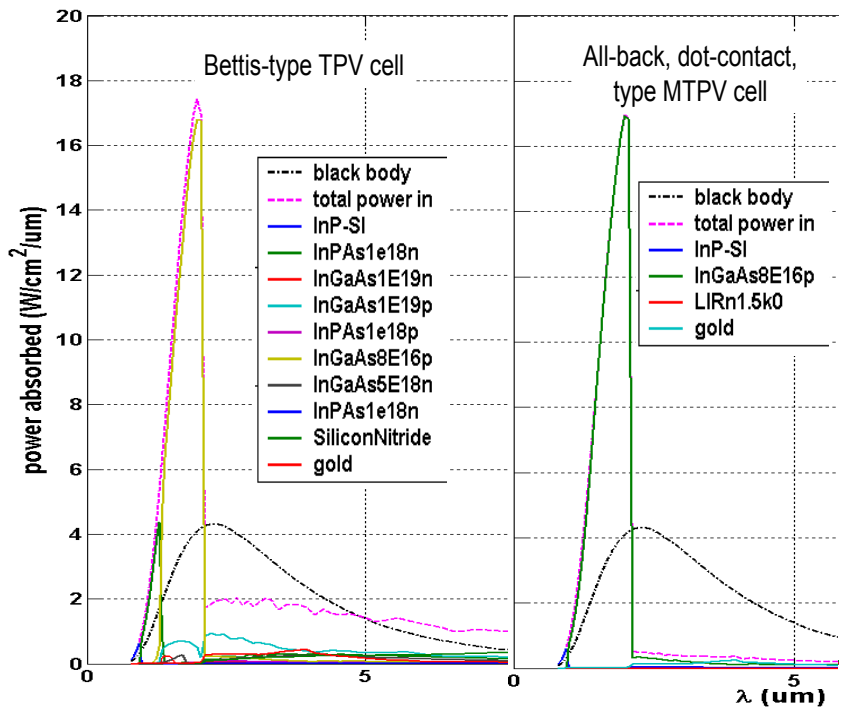

Fig. 3. Comparison of optical absorption characteristics for a >20% efficient TPV cell and a MTPV dot-contact cell with a 1000°C greybody-emitter spectrum and 0.1 μm gap.

**MTPV emitter**

Just as the cell for MTPV operation must be especially designed for the application, so must the emitter. Many of the useful TPV filters and structures interfere with or vitiate the power enhancement provided by MTPV. For example, normal spectrally-selective, anti-reflecting (AR), or reflecting thin-film coatings do not work well over the extremely-wide range of light-transfer angles in the microgap structure. (As the gap gets smaller, the critical angle gets larger, and more of the emitted light goes through the coatings at non-optimal angles.) Furthermore, few of these coatings have the high average refractive index required for utilizing the $n^2$ effect. (Nevertheless, high-index AR coatings have been explored extensively for specific efficiency benefits.)[14] The emitter coatings modeled below act as AR coatings to some extent (some showing peak optical power output as a function of thickness at specific frequencies), but it is believed that a combination of other things contributes greatly to, or dominates, this effect. For example, a resonance coupling across the gap will reduce the energy reflected at the surface. However, that has nothing to do with the standard AR coating model, which has coherent-light interfering at a surface to suppress reflection.

No conventional AR coatings have the appropriate resonance properties to go beyond the $n^2$ effect, since they are chosen to avoid absorption and their thickness actually increases the effective gap between the high refractive-index materials. On the other hand, with the proper materials, the thickness of even very-thin layers can have a significant impact on emitter output and spectral selectivity (see below). The $n^2$ effect is a two-edged sword. While increasing the refractive index of the emitter and detector increases the energy available for transfer, it also reduces the optical coupling into the gap. Unless the gap is reduced significantly below the λ/2 range, the coupling losses dominate the access to internal-energy gain (Fig.1).

Early-on in the project, recognition of a fundamental property of near-field coupling almost became a "show stopper" for MTPV. Since power-transfer enhancement from the near-field effect across a gap d, is a function of the ratio (λ/d) in the region of interest, the useless and harmful long-wave light is enhanced more than is the useful light. Even with the useable items of proposed filters and the demonstrated high-efficiency TPV cells (as seen in Fig. 3), MTPV power output might be good, but its overall efficiency would be poor. The need for high power output and high efficiency, both within the constraints of the sub-micron gap and non-contacting requirements of the MTPV structure, led to a major effort to find materials that could utilize and optimize the $n^2$ and the resonance effects.

As a starting point, semiconductors, with their high refractive index and band-gaps (with low spectral absorption – and emissivity) matching those of the PV detector were a natural choice for the emitter. Using silicon as an example,[11] it became obvious that, because of free-carrier absorption at high temperatures, very-thin layers were required. Similarly, direct-bandgap semiconductors with their more-abrupt and higher absorption at the bandedge would provide the best results.

**COMPUTATIONAL ANALYSIS**

The computational analysis model (developed at MIT and Draper Labs)[15,16] is now a particularly useful tool because it allows many different materials to be included in various layers. It is possible to model actual devices based on theoretical or, better yet, on experimentally-determined material parameters. The limitation of the emitter model is that experimental high-temperature optical parameters are not available for many useful materials. Thus, estimates had to be made for the emitter parameters. Nevertheless, some amazing confirmations and predictions can be made clear with this tool.

The effect of the near-field coupling on power transfer between an emitter and a greybody detector is indicated in Fig. 4. This shows the spectral power from a semi-conductor emitter (silicon) and its useful improvement with a single-emitter coating of varying thickness. The most important such coating is a modification to a Drude filter[10,17,18] that suppresses IR transmission from the emitter bulk. Even a 10 nm thickness is sufficient to reduce the long-wave emission (across a 100 nm gap) by an order of magnitude. The figure indicates how increasing the single-coating thickness further can change the system from high-power and modest efficiency to one with 2/3 the power-output, but with double the efficiency. This large "trade-space" is important in terms of the number of applications for an MTPV system. For example, terrestrial applications (with fuel and cooling readily available) would find the high power output most appealing. Whereas, space applications, for which cooling and input power are limited, would require the high-efficiency system.

Comparison can be made between Fig. 4 (a spectrally-selective emitter and a matching, high-refractive-index, greybody detector) with Fig. 3 (a high-$n$, greybody emitter and spectrally-selective TPV detectors). Both show the $n^2$-type enhancement in the useful bandgap region. Both show the greatly-reduced, long-wavelength absorption (produced by different techniques) that would otherwise ruin MTPV efficiency. Neither alone is sufficient to produce an optimal MTPV system. Fig. 4 also shows a resonance spike at the selected bandgap as a result of surface effects of the selective emitter (not seen in Fig. 3). This effect will be further developed in a later section.

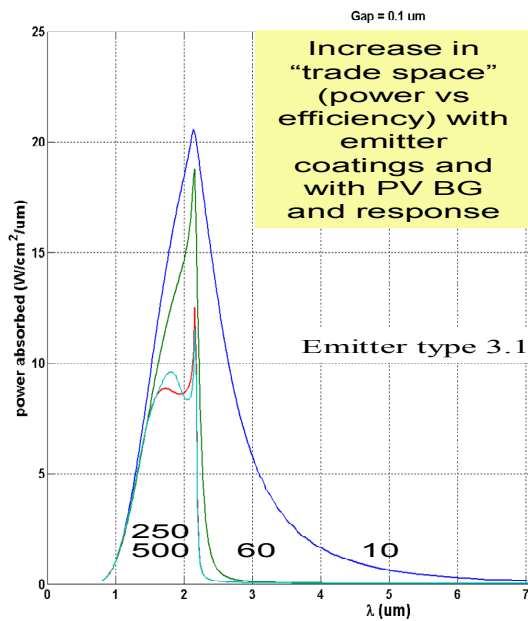

Fig. 4. Effect of emitter IR-suppression-coating thickness (10-500 nm), tuned for a 2.2 micron bandgap device, on spectral-power transfer from selective emitter into grey body across a 100 nm gap.

Coatings developed for the MTPV application are often beneficial for TPV products as well. The spectral emissivity (spectral power output relative to black-body output) of an emitter at a given temperature (e.g. 1000°C) and at two gap distances (0.1 and 100 μm) is seen in Fig. 5 to vary by more than a factor of 5 in the useful wavelength region. This calculated output-power enhancement is primarily a result of the $n^2$ effect; but, other effects are identified as well. The semiconductor-emitter coating layer must be thin, so that free-carrier absorption/emission at high temperatures does not introduce useless long-wave light.[11] Use of a semiconductor as an emitter as well as a detector is a form of resonance coupling. The hot semiconductor emitter is chosen to have the same band edge as the cold PV device. Thus, the high absorption in the emitter becomes high emissivity to couple into the PV-device active region. This is a macroscopic effect. However, as to be seen, it can become a microscopic resonance effect leading to significant enhancement of the optical coupling.

Surface-plasmon-coupling spikes are identified in the narrow-gap configuration. However, since the model integrates over all angles, these directional spikes are not as dramatic as when viewed at their characteristic angles. These are local resonances (i.e., not seen in the far-field picture of Fig. 5, where "noise" may be seen from interference with reflections between the high-n surfaces).

The selected MTPV spectrally-selective-coating thickness is much more important for near-field coupling than for far-field emission. In both cases, the useless IR radiation is suppressed. However, the near-field coupling of useful light is more strongly affected. The high-refractive-index emitter coatings in the example, required for optimal near-field coupling, are not optimal as AR coatings for far-field emission. (Tuned AR coatings on <u>both</u> emitter and detector can give an emissivity of 1 for excellent far-field results.)

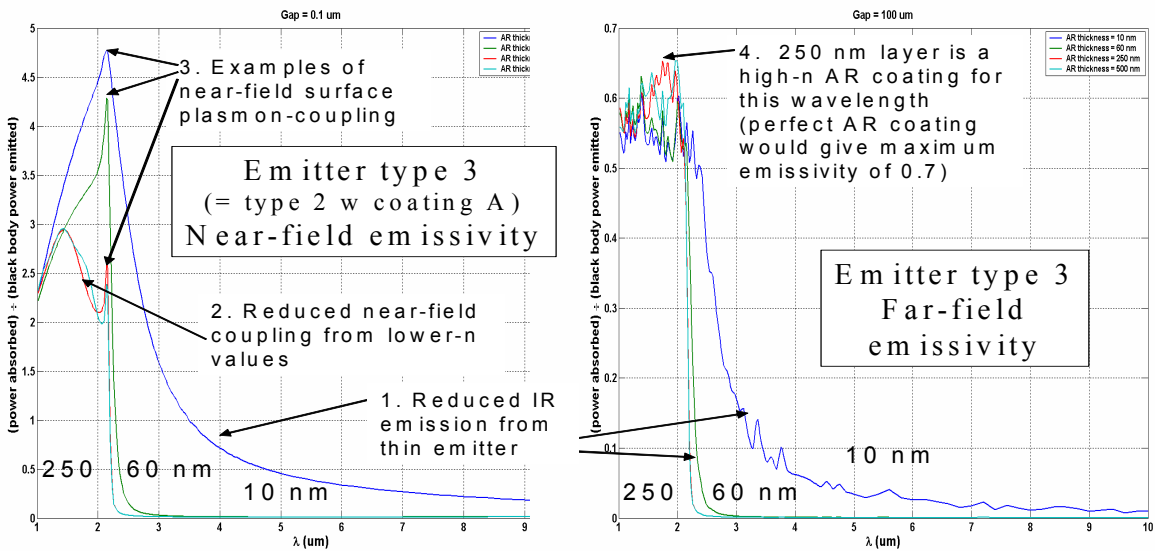

*Fig. 5. Comparison of near-field (0.1 μm gap) and far-field (100 μm gap) emitter/detector spectral-power transfer (in terms of emissivity). Four emitter-coating thicknesses are compared (10, 40, 250, and 500 μm)*

With this type of emitter/TPV detector combination, the near-field spectral efficiency (useable light energy absorbed divided by total light energy absorbed by the detector) can exceed 90%. With optimized MTPV devices (based on actual TPV devices, but modified for MTPV operation), thermal-to-electrical power conversion efficiencies in excess of 30% are predicted at emitter operating temperatures of ~900°C. Fig. 6 explores the modeled development of, and trade-space available to, the MTPV system. The power output and cell efficiency is pictured for the two cell types in Fig. 3 as a function of changes in cell and emitter properties (material, design, and layer structure).

The first dataset in Fig. 6 shows that while the electrical <u>power output</u> of the two cell types do not differ greatly in the MTPV operational mode, the <u>efficiency</u> of the MTPV cell is 2.5x that of the TPV cell. The cell efficiency in this figure includes the losses due to optical coupling (absorption of non-useful light into the cell). In the second dataset, with an appropriate IR-blocking layer on the emitter, the cell powers don't change, but their efficiencies increase by >10% and 50% for the MTPV and TPV cells respectively. The TPV cell efficiency improves so much because it was not as strictly designed to reduce IR absorption as was the MTPV cell. If efficiency is more important than power output, increasing the thickness of the IR-filter layer (see Fig. 4 and cluster 3 of Fig. 6) raises the efficiency, but at the cost of cell output power. Additional selective-emitter coatings can improve the optical coupling in the useful range (cluster 4 of Fig. 6). Continuation of this partially-optimized process (in the 5$^{th}$ group of Fig.6) results in close to the unfiltered power with close to maximum cell efficiency.

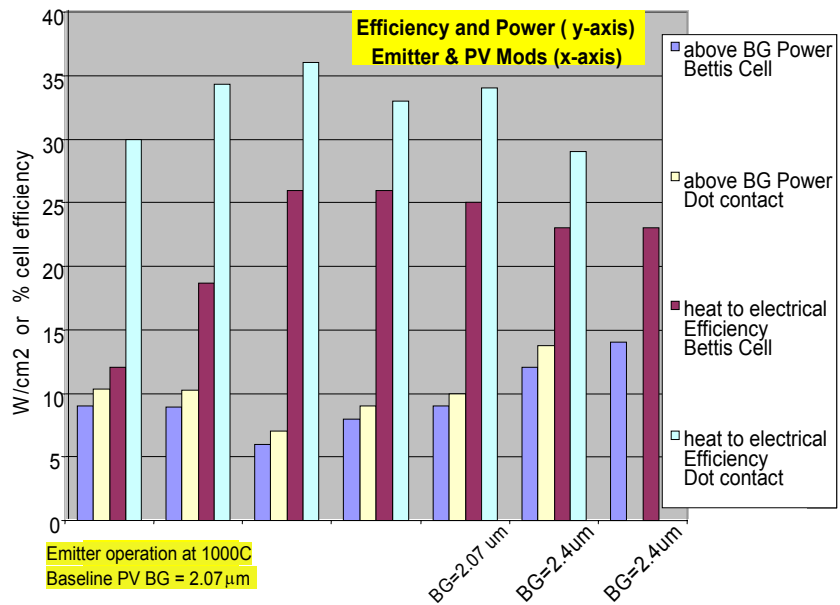

Fig. 6. Power output and cell efficiency for two cell types as a function of changes in cell and emitter properties (see text).

The last two groups of the figure show the results for a change not previously discussed in this paper, the MTPV device bandgap optimization. By increasing the semiconductor bandgap from 2.07 μm to 2.4 μm, it is possible to increase the cell output power by ~40%. There are penalties in efficiency to be paid for this gain in power (e.g., lower cell voltage and therefore more photons with energy wasted above bandgap). The last group shows how, since the TPV cell already incorporates heavily-doped layers, it is possible to improve its power output (e.g., by going to a Tandem-cell structure) without significantly lowering the efficiency. Such a change in the MTPV cell design would have a major negative impact on the system efficiency and therefore is not shown.

This figure indicates the two major thrusts of the MTPV project: improvement of the detector and improvement of the emitter. With unmodified emitters, the MTPV cell design is critical to improved efficiency. However, with the MTPV cell design, improvement in the emitter design, while significant, may not be worth the price (cost-wise, stability-wise, or operationally). However, the TPV cell exists; therefore MTPV-operating efficiency can be doubled by development of the selective-wavelength emitter without having to develop a new cell. If the selective filter is found to have unexpected problems, then development of the MTPV cell provides the option for more than doubling the TPV cell efficiency. Of course, having both improvements is best, since the full trade space and range of applications is then available.

**Compound Resonance**

We have mentioned the issue of resonant coupling from the emitter (to improve selective emissivity) and from the detector (e.g., with quantum wells) side of the system. We have not yet gone the final (?) step. Figure 7 demonstrates the effect of compound near-field resonant coupling (e.g., when both the detector and emitter have "tuned" thin layers with a common oscillator frequency across a 0.1 μm gap). The emitter is the same in both systems pictured; only the detector characteristic is changed.

An order-of-magnitude enhancement of the compound-resonant energy (only the spike at the band edge) is seen with the addition of a resonant layer (optimal thickness < 250μm) to the detector. [The power in the spike without the detector resonator is about ½ of the peak height or ~6 W/cm$^2$/μm.] Such compound-resonance effects have been reported in the past,[19] pursued theoretically,[17,20] and even new devices[21] have been proposed to benefit from them. However, this option does not seem to have been followed up with a practical implementation, yet. It is not known whether saturation of the effect is observed at 100 - 200 μm because that matches the gap width or because of the material parameters chosen. Comparison of the compound-resonant and resonant detectors clearly indicates some additional interference effects between the thin layers that could be further exploited.

An observation that needs further study is that resonance enhancement from this doubly-resonant structure (relative to the other sources) is comparable to that seen for a singly-resonant structure at a much smaller gap (10 nm).[17] The power transfer is not comparable (100 W/m² vs. 2x10⁶ W/m²), since that has the strong dependence on spectral selectivity, gap thickness, and emitter temperature. However, this strong enhancement of spectrally-selected wavelengths is very important to cell and system efficiencies and it is available at the attainable gap thicknesses where the $n^2$ effect is otherwise dominant. In the authors' opinion, this compound-resonance enhancement has more practical near-term potential (for large-area devices) than do reliable 50 nm gaps.

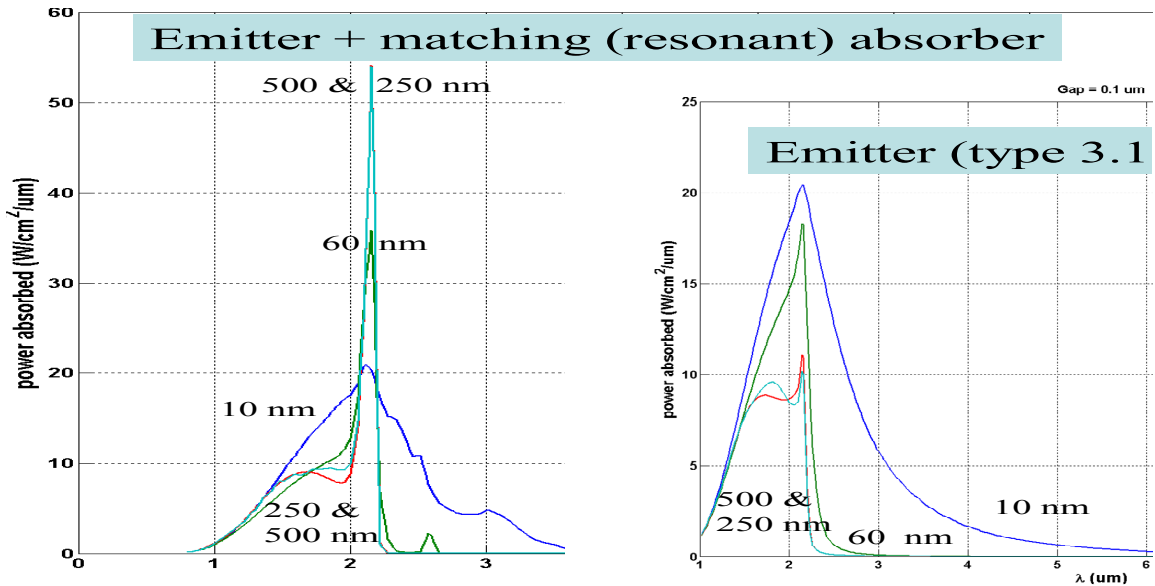

*Fig. 7. Comparison of near-field (0.1 μm gap) emitter/detector spectral-power transfer when the greybody detector is changed to contain a matching (resonant) surface layer. Four resonant-coating thicknesses (equal on both emitter and detector) are compared (10, 60, 250, and 500nm)*

Conventional absorber materials of the stated thicknesses were used in this model (for Fig.7), rather than the thinner, high-Q, resonant quantum-well or quantum-dot structures. The surface plasmon effect was used to simulate the quantum-well resonance. However, this beyond-$n^2$ enhancement and total power transfer in the figure is only about 20% of that predicted by quantum-mechanical calculations for quantum-well to quantum-well structures at a much-lower temperature.[3] The difference might be accounted for by a strong angular dependence of the surface effects that does not limit the quantum-well results in a similar manner. Nevertheless, the predicted quantum-well resonance effects are seen qualitatively in the enhancement of a layered-emitter simulation. For example, resonant spectral-control-coating thickness (similar to that of the quantum well) is seen to be very important – up to a point (e.g., when coating thickness approaches gap width - and the source material is thus moved further from the absorber material). A different dependence on QW width, and on distance between the resonators, is expected, since the effective electric-dipole moment of the quantum-well is much larger than that of atoms in the selective filter and the resonance frequency changes with QW width.

The materials and structures represented here were optimized over a multi-year development among several laboratories for both the spectrally-selective emitters and TPV devices, but not for the compound-resonant detectors. Some of the concepts had been developed decades earlier.[14] Some aspects of this development are still proprietary. Experimental work to confirm the modeled resonant coupling was never developed beyond the design stage[4], in part, because the theoretical and computational predictions were in sufficient agreement and the results were so compelling and, in part, because the actual values of the material optical properties at high temperatures were not available. Without confirmation of these input parameters, further optimization and even-higher efficiency or power-conversion values would not be any more convincing than those results already presented here.

Front-surface quantum-well or quantum-dot photodetectors, with a characteristic operating frequency, can be matched to the emitter resonance to create an efficiency-optimized system. The total power output might be less than for an un-optimized system (because of the small volume of the quantum structures relative to a macroscopic layer); but, enhancement of light coupling at highly-resonant, relative to non-resonant, frequencies would greatly reduce conversion losses if resonance is close to the device bandgap or to the quantum-structure absorption resonance. In the context of Fig. 6, MTPV cell efficiencies on the order of 50% could be expected. (The gap width here is too large to allow coupling-induced radiative recombination to reduce MTPV-cell performance.)[17] Therefore, both the cell and system efficiency would be very high. The trade-off between power output and system efficiency would determine the structure dimensions and materials chosen for a particular application. It is clear that the "trade space" is quite large for such systems.

**DISCUSSION**

Sponsorship of this work was terminated over five years ago (with the expectation that, even without the high-temperature-material optical properties necessary to confirm the modeled results, it was ready for commercialization). Since then, a significant body of work elsewhere has addressed optical power transport across the sub-100 nm gap range. However, very little work has been published on practical devices for the near term. One exception is reference 17 (Larouche, Carminati, and Greffet); and even that paper pushes into the multi-nanometer-gap range, rather than optimizing the more practical tenth-micron gap range that we address. Nevertheless, since there is an overlap at 100nm, some comparisons between that and the present work can be useful (Appendix B). This comparison is very important in identifying some of the issues spelled out in this paper and therefore in guiding future modeling and experimental work.

The ability of the multilayer model to accurately model interaction, emission, and absorption, of thin layers of different materials and the reflection of non-absorbed light from the different detector layers back into the emitter is critical to the development of a successful MTPV system. Nevertheless, it is still only possible to simulate and not to accurately model the quantum structures that can create the truly strong resonances that are foreseen by some models.[3,21] A few structures and materials have been optimized using the Draper Code. However, there are new ones awaiting exploration and integration into the new capability. High-temperature material parameters (optical and electrical) need to be determined, and many that would not have survived the high temperatures of prior TPV requirements are now candidates for inclusion as parts of selective emitters in the 700 - 1000°C range. This is a wide open field with immediate and extensive applications and great variety in approaches to them.

**CONCLUDING REMARKS**

Computational capabilities have confirmed the resonant-proximity effects of both $n^2$ and "beyond $n^2$" enhancement, thereby removing a major limitation in ThermoPhotoVoltaics. These capabilities, which model thin layers of multiple materials, have demonstrated the increased efficiency of TPV converters from highly-selective emissivity. Furthermore, they show the improved emitter radiative efficiency from return of energy that would otherwise be lost from the emitter or in the PV converter (in the form of radiated long-wavelength light or heat absorbed). This formerly lost energy is recycled into higher frequencies, selectively emitted at more appropriate wavelengths, and even coupled into a compound-resonant MTPV converter.

A major purpose of this paper is to address the efficiency issue of MTPV, not to maximize the energy transfer. The spectral selectivity of the emitter and detector, required to attain high efficiency, lowers total electrical power output. Therefore, an increase of power across the micro-gap at all frequencies makes up (at least to some extent) for the reduced power output from spectrally-selective emitters. As a consequence, in none of the figures presented (all with 100 nm gaps), is the $n^2$ times black body value ever reached (even for the compound resonance example). The "beyond $n^2$" effects can be seen and are very effective, even at gap widths as great as 100 nm, but they may not push the $n^2$ limits until the gap is reduced further. This points to the margin-for-growth (without technological breakthroughs), that is so important for commercial viability.

A key feature to remember is that the new energy-transfer mechanism does not depend only on the release of the blackbody radiation trapped within the emitter (via the classical $n^2$ effect). Additional energy-transfer sources are the non-propagating photon modes (near-field radiation normally dissipated in self-excitation of the emitter atoms) and the resonance-coupling effects.

Despite the availability of this extra energy, the blackbody law of power emission (which pertains only to the propagating-photon modes) and conservation of energy are not violated. We can <u>not</u> continuously get more power out than we put in. However, we can extract energy more rapidly and more selectively at any given emitter temperature. While this benefit might appear to be coming from Maxwell's demon[22] working overtime, it is actually a consequence of recognized material properties, their interactions, and the ability to establish resonances within and between them. Therefore, with microgap coupling and a given thermal-energy input, the emitter can be kept at a lower temperature and still operate at a higher power and efficiency than previously possible. This lower-temperature operation is important in that it opens the possibility of selecting emitter materials and structures that would not survive higher temperatures. It also improves the efficiency of heating the emitter with a given source. The end result is a practical MTPV system with many thermal-to-electrical energy-conversion applications, from high-end space-power systems to devices for getting electricity from solar furnaces, gas lanterns, or even a small cooking fire.

**ACKNOWLEDGEMENT**

Part of this paper is based on work at Draper Laboratory, Inc., Cambridge, MA 02139 in 2004 and 2005. Subsequent work was funded in part by HiPi Consulting, New Market, MD, Science for Humanity Trust, Bangalore, 560012, India and Science for Humanity Trust, Inc, Tucker, GA, USA. Special thanks are due Drs. Mark Weinberg and Eric Brown of Draper Labs respectively for setting up the computational model used here and for extending its capabilities and for making it so user friendly.

**APPENDIX A: SELECTIVE SPECTRAL EMISSIVITY**

Figure A-1 (the same results as in Fig. 3, but in a semi-log plot) indicates the precision and detail available for modeling the various PV-cell and emitter structures to be studied. This model is representative of a particular cell type (power-absorption response shown for MTPV operation with a high-$n$, greybody-emitter spectrum) that has high efficiency in a TPV operating mode (>20%). Nevertheless, it is only a mediocre performer (>12%) in an MTPV mode with a similar emitter.

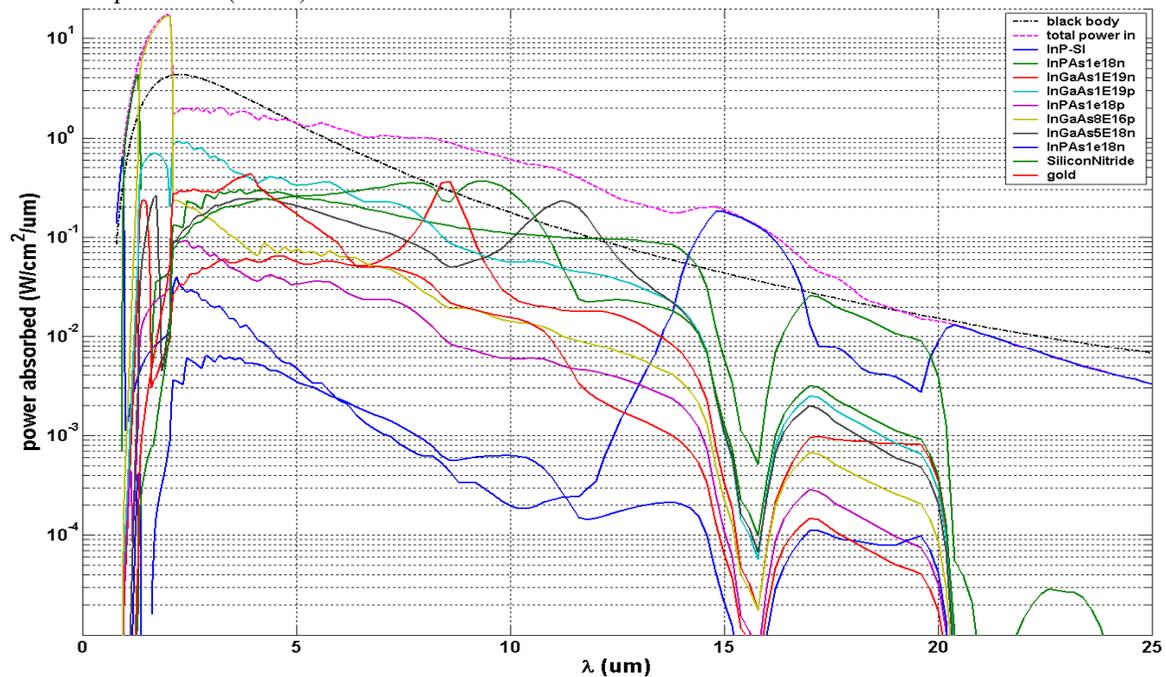

*Fig. A-1 Modeled optical power absorption within a 20% TPV cell operating under MTPV conditions with a greybody emitter (blackbody curve shown). These are the same results as in Fig. 3, but in a semi-log plot.*

The benefits of the $n^2$ proximity effect are seen in the higher-than-blackbody power absorption below the band-edge wavelength (in the active PV layers). But, the effects of heavy-doping are seen in the high absorption of the numerous thin n and p layers. (These include the absorption peaks associated with the heavy-doping effects and thus the plasma resonances of the Drude model.) While individually these layers are not very absorbing, the fact that there are so many of them is a problem. Their cumulative effect, along with the long-wavelength enhancement, makes the total absorption of useless light significant. From such calculations, it is possible to optimize the emitter / PV device system (sometimes one layer at a time, but generally by following the recognized rules expressed in Fig. 2).

**APPENDIX B: DETAILED COMPARISON OF THIS PRESENTATION WITH REFERENCE 17**

Since there is an overlap in the analyses at 100nm, some comparisons between Reference 17 and the work presented here might be useful. The similarity between the reference's Fig.4 and our Fig.1, showing the expected power gain as a function of decreasing gap thickness, indicates the common basis for a practical interest in this field.

Another common feature, identified above, is the use of a Drude emitter/filter to suppress emission of useless long-wavelength light. However, reference 17 simulates the Drude emitter as a quasi-monochromatic source and we have selected a real material (by extrapolating its optical properties to high temperatures, e.g., ~1000°C). Therefore, neither paper has real material values to work with. Nevertheless, this is a critical feature of both models. The reference figures (3 and 5) indicate something that we discovered early in our exploration with the model; a Drude material by itself is not adequate to give the spectrally selective results that we seek. The tungsten emitter of the reference (their Fig. 3) is actually better than their Drude emitter (Fig.5) at gaps larger than 30 nm. Those two figures do show the same increase in optical-power transfer with decreasing gap width that we see in our Figure 5. However, the improvement of useful power in our Figure 5 is greater than that shown for the same gap width in Figure 3 of the reference because we have better matched the high refractive index of both emitter and detector. (The tungsten emitter in the reference cannot have the high refractive index of the detector in the wavelength region of interest.) This demonstrates that the $n^2$ effect in this gap width and wavelength region can be more important than the surface-resonance effects.

Other ways in which reference 17 differs from our work is the display of the results (rad/s vs. wavelength on the abscissa), the emitter temperature (2000K vs. ~1000C), and the modeling of the TPV cell and emitter. The reversal of the horizontal display in their Fig. 5 makes their result look like our Figs. 3, 5, and 7, when in fact they are quite different. Their figure shows a significant IR (low frequency) content whereas ours do not. It also shows a spike in the high-frequency (short-wavelength) region (near $2 \times 10^{15}$ rad/s or ~1.4 eV) whereas we have a spike at 2 microns (~ 0.6 eV). This is a result of each group optimizing the Drude filter in a different region. Their higher emitter temperature, means that they have a lot of energy in a region of the emitter spectrum where we have nearly none. So they have tuned the filter to maximize their output. Since our thrust is to maximize efficiency, we have tuned it close to the bandgap of our detector. The result is that they have a broad peak on the origin side of the spike (as we do), but it means that they are absorbing non-useful long-wavelength energy in their detector.

Larouche, Carminati, and Greffet mention use of a 300 μm GaSb slab as the cell (although their paper does not indicate whether it is optically modeled by a slab with air behind it or by a semi-infinite slab - even this can make a difference). We used a 10-layer cell model, fully representative of a real TPV cell, or a 4-layer model, representative of our proposed MTPV cell. As a consequence, figures 3 and 5 of the reference show strong radiative-power exchange in the far IR region, which we have prevented by emitter and cell design.

Our Figure A-1 indicates the importance of having multilayer capability to guide that design. I believe that reference 17 modeled a single-layer emitter, whereas we used a 3-5 layer emitter. Our Figures 4, 5, and 7 indicate the importance of thickness in the emitter layers. To simulate the improvements possible, beyond the IR suppression from a modified Drude filter, we added an additional surface layer (Fig. B-1) to increase the $n^2$ effect. This is possible because the Drude filter, while critical for the efficiency of a TPV or MTPV system, does not have the high refractive index of a semiconductor detector in the wavelength region required. Without having to resort to photonic crystals, we can make a significant improvement in the emitter-detector coupling by the addition of a very-thin layer of high-refractive index material to help match the effective index of the emitter with that of the detector. In a sense, it is an antireflective coating, just not of the interference type. In another sense, it is a layer designed to create a cross-gap or compound resonance.

Notice that even a 60 nm layer increases the power transfer of useful wavelengths by 20% (Fig. B-1). However, there is a cost in terms of system efficiency since the IR transport is also increased. This is the mechanism responsible for the improvement seen in going from group 3 to group 4 in Figure 6. Two other interesting items can be seen in the figure. First, even the thin layer of this material is sufficient to eliminate the surface resonance (the spike above 2 μm disappears between 10 and 60 nm). Second, the curves with a thicker coating begin to show the actual shape (not the apparent shape) seen in Figs. 3 and 5 of ref. 17.

What do these differences in the model mean? We obtained a lot of information from our early 5-layer model (typically 3 layers for the emitter, 1 for the gap, and 1 for the cell - as in reference 14), which we used for more than a year. We could even have confirmed our present design for suppressing long-wave IR coupling with the Drude filter, had we thought of it at the time. However, we were not able to make real progress in the areas of predicting benefits of a new MTPV cell design or of optimizing the emitter design until we could go beyond this limitation. With the capability of a many-layer code, which allowed us to properly model real cells and to develop truly spectrally-selective high-refractive-index emitters, we obtained TPV-cell efficiency vs. gap-width curves similar to Figure 12 in the reference (shape and values). However, our present heat-to-electricity efficiency for a modeled TPV cell (at ~ 25%, Fig. 6) were obtained with a much lower temperature emitter and at near-future-attainable gap widths (0.1μm). We also display much higher efficiency results (up to 35%) for the specially-designed MTPV cell under these realistic conditions. Both cells have modeled power output in the multi-MW/m² range with 1000C emitters and 0.1 μm gaps. Reference 17 obtained those same power levels with a 2000K Drude emitter and with a gap below 10 nm.

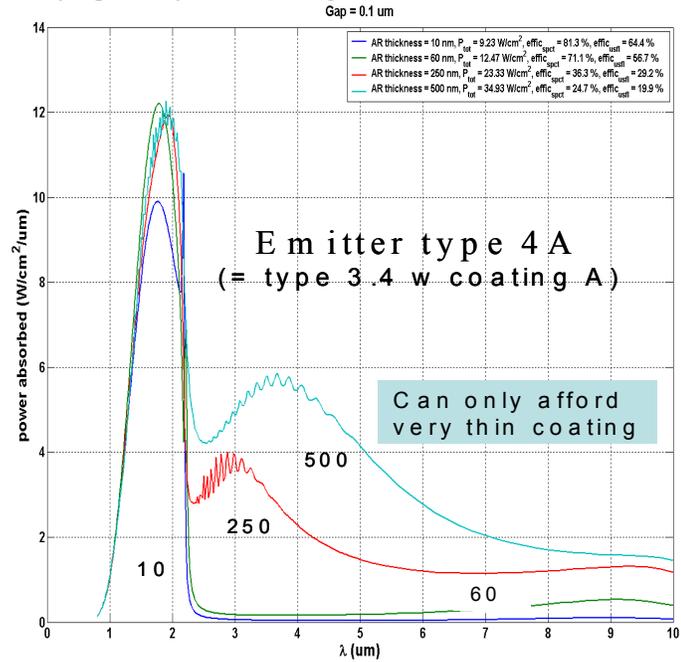

*Fig. B-1. Effects of adding a high refractive index coating (10 to 500 nm thick), to the same single layer coating displayed in Figs. 4, 5, and 7.*

In our discussion of the compound resonance, we mentioned that our 0.1 μm gap, with mutually resonant layers on either side (Fig.7 - left side), gives resonance-enhancement results similar to those in Reference 17 for a 10 nm gap (their Fig 5a, with only a Drude emitter and no matching resonator in the detector). Furthermore, some of the curves on the right side of our Fig. 7 (without a matching resonator) appear to be the same as their Fig. 5b for a 30 nm gap. This false similarity was explained above as a result of: the higher temperature used in the reference (thereby increasing the higher-frequency light emission); the use of thicker detector/absorber (300 μm vs. our ~6 μm of useful absorbing region), which gives higher IR (low frequency) absorption in the reference work; and the higher conductivity of their Drude filter (thus a higher plasma frequency and resonance peak needed to optimize their system for the higher temperature) compared to ours.